%% file: adce_arxiv.tex
\documentclass[conference]{IEEEtran}

\usepackage{amsmath,graphicx}
\usepackage{nicefrac,xfrac}
\usepackage{lipsum}
\usepackage{amsmath,amssymb,amsfonts,accents,mathrsfs}

\def\bxi{{\boldsymbol{\xi}}}
\def\bH{{\mathbf{H}}}
\def\bh{{\mathbf{h}}}
\def\bY{{\mathbf{Y}}}

\def\bS{{\mathbf{S}}}

\title{\huge Study of Joint Activity Detection and Channel Estimation Based on Message Passing with RBP Scheduling for MTC \vspace{-0.5em}}

\author{\IEEEauthorblockN{Roberto B. Di Renna and Rodrigo C. de Lamare}
\IEEEauthorblockA{Center for Telecommunications Studies (CETUC)\\
Pontifical Catholic University of Rio de Janeiro, RJ, Brazil\\
Email: \{robertobrauer,delamare\}@cetuc.puc-rio.br}
}

\begin{document}
\maketitle

\begin{abstract}
    In this letter, based on the hybrid generalized approximate message passing (HyGAMP) algorithm, we propose the message-scheduling GAMP (MSGAMP) algorithm in order to address the problem of joint active device detection and channel estimation in an uplink grant-free massive MIMO system scenario. In MSGAMP, we apply three different scheduling techniques based on the Residual Belief Propagation (RBP) in which messages are generated using the latest available information. With a much lower computational cost than the state-of-the-art algorithms, MSGAMP-type schemes exhibits good performance in terms of activity error rate and normalized mean squared error, requiring a small number of iterations for convergence. %
 \end{abstract}

\IEEEpeerreviewmaketitle

\section{Introduction}
\label{sec:intro}

    Differently from the conventional human-type communications, massive machine-type communications (mMTC) for the Internet of Things (IoT) have unique service features, such as uplink traffic dominated by short packets~\cite{DiRennaAccess2020}, sparse device activity~\cite{Salam2019}, high energy efficiency~\cite{Fuqaha2015}, low data rates~\cite{3GPP2019} and uncoordinated access~\cite{MShirvanimoghaddam2017}. Since the total number of machine-type devices (MTDs) is much larger than the receive processing resources, conventional scheduling-based orthogonal multiple access schemes are not suitable. Due to the high probability of frame collisions, the scheme that the base station (BS) allocates orthogonal time/frequency resources to each device is impractical for the mMTC scenario. Moreover, sporadic short packets in the mMTC traffic are compromised due to signalling overhead and excessive latency. The uncoordinated access proposed in recent years can solve those issues. Based on grant-free \textit{non-orthogonal multiple access} (NOMA)~\cite{MShirvanimoghaddam2017}, active devices transmit frames without previous scheduling, in order to eliminate the need for round-trip signaling. Thus, it is up to the BS to estimate the channels, detect active devices and transmitted signals.

    Several approaches have been considered for joint user activity and data detection ~\cite{Zhu2011, Knoop2013, DiRennaWCL2019,DiRennaTCom2020}. In most of these studies, the uplink channel state information (CSI) from the MTD to the BS is assumed to be perfectly known to the BS, which allows interference cancellation techniques \cite{itic,DeLamare2008,jidf,rrser,mfsic,mbdf,tds,bfidd,aaidd,listmtc,1bitidd,detmtc,dynovs}. However, in practice, the uplink CSI should be estimated before data detection. Since the channel vector is also sparse, compressed sensing (CS)-based technique is a good fit for the scenario. In this way, the multi-user detection (MUD) problem can be seen as a sparse signal recovery problem. Exploiting the \textit{a priori} distribution of the sparse vector to be recovered, the works in~\cite{LLiuTSigPr2018,ZChenTWirCom2019,KSenel2018} present denoising-based AMP algorithms verifying the activity error rate performance. As an extension of the GAMP~\cite{Rangan2012} algorithm, the inference algorithm HyGAMP~\cite{Rangan2017} is then developed. Since the components of the channel are particularly independent with conditional distribution, the combination of a loopy belief propagation (LBP) part for user activity detection and a GAMP-type strategy for channel estimation makes HyGAMP outperform other existing algorithms in terms of mean square error (MSE). However, HyGAMP considers a completely parallel update of the messages, where each iteration performs exactly one update of all edges.

    Sequential scheduling significantly improves the performance in terms of convergence and error rates~\cite{CasadoTCom2010}. The main idea is to find the best sequence of message updates, focusing on the part of the graph that has not converged. The residual belief propagation (RBP) is an informed dynamic strategy that updates messages according to an ordering metric named residual. The residual is the difference between the messages of the actual iteration and the previous. Thus, in order to find a more efficient implementation or/with better convergence solution, in this work, we propose a message-scheduling GAMP (MSGAMP) algorithm along with three message scheduling strategies. The proposed MSGAMP algorithm and strategies exploit the \textit{a priori} distribution of the sparse channel matrix and use the number of antennas in the base station to improve the activity detection. Simulations show that MSGAMP results in an improved performance over HyGAMP in terms of NMSE while requiring a small number of iterations for convergence and a lower computational cost than HyGAMP.

    This paper is structured as follows. Section II introduces the system model. The problem of joint channel and user activity estimation along with the proposed MSGAMP message passing algorithms is detailed in Section III. Section IV presents and discussed the results of simulations, whereas the conclusions are drawn in Section V.

% System model
% -------------------------------------------------------------------------
\section{System model}
\label{sec:sys_mod}

    We consider the grant-free uplink NOMA scenario. Assuming there are $N$ single-antenna devices communicating with a BS equipped with $M$ antennas \cite{mmimo,wence}, the problem of interest here is to estimate the channel matrix $\mathbf{H} \in \mathbb{C}^{N\times M}$ from a received signal $\mathbf{Y}$. Considering $\mathbf{a}_n$ as independent pilot sequences, the received signals are obtained through the following model
    \begin{align}\label{eq:sig_model}
        \mathbf{Y} &= \sum_{n=1}^{N} \frac{\boldsymbol{\phi}_n}{\|\boldsymbol{\phi}_n\|} \mathbf{h}^T_n + \mathbf{W} = \boldsymbol{\Phi}\, \mathbf{H} + \mathbf{W},
    \end{align}

    \noindent where $\mathbf{W} \in \mathbb{C}^{L\times M}$ is the independent complex-Gaussian noise matrix with $\mathcal{C}\mathcal{N}\left(0,\sigma_w^2\right)$. The pilot matrix $\boldsymbol{\Phi} \in \mathbb{C}^{L\times N}$ is composed by $\boldsymbol{\phi}_n = \exp{\left(j \pi \boldsymbol{\kappa}\right)}$ of each device, where each element of vector $\boldsymbol{\kappa} \in \mathbb{R}^{L}$ is drawn uniformly at random in $\left[-1,1\right]$ and $L$ is the length of the pilot sequence. Each active device transmits $L$ pilot symbols, which we denote here as a frame. Since the frame size of each device is typically very small, we assume that all devices are synchronized in time.

    Each element of $\mathbf{H}$ represents the channel gain between the $n$-th device and the $m$-th BS antenna. Since mMTC is a sparse scenario, we denote the Boolean variable $\xi_n \in \left\{0, 1\right\}$ that indicates if the device is active when $\xi_n = 1$ and inactive otherwise. Thus, considering as $\rho_n$ the probability of being active of the $n$-th device, $P\left(\xi_n = 1\right) = 1 - P\left(\xi_n = 0\right) = \rho_n,$ where all $\xi_n$ are considered i.i.d. and each device has its own activity probability. Therefore, it is possible to define the the prior distribution of $h_{nm}$,
    %\vspace{-5pt}
    \begin{align}
        P\left(h_{nm}\right) &= \left(1-\rho_n\right)\delta\left(h_{nm}\right) + \rho_n\, \mathcal{C}\mathcal{N}\left(h_{nm} | 0,\beta_n\right),
    \end{align}
    %Thus, given the vector $\boldsymbol{\xi}$, the components of $\bH$ are independent with conditional densities given by
    %\begin{equation}\label{eq:rand_var_xi}
    %    h_{nm}|\boldsymbol{\xi} \sim \left\{
    %    \begin{array}{ll}
    %        \delta{\left(h_{nm}\right)},                                        & \xi_n = 0, \\
    %        \mathcal{C}\mathcal{N}\left(h_{nm} | 0,\beta_n\right), & \xi_n = 1,
    %    \end{array}\right.
    %\end{equation}
    \noindent where $\delta\left(\cdot\right)$ is Dirac delta function and $\beta_n$ is the channel variance of each device. As in the mMTC scenario $N$ is larger than $M$, the system is overloaded. However, due to the low activity probability of devices, $\mathbf{H}$ is sparse, which makes its recovery possible through the theory of compressed sensing (CS)~\cite{JWChoi2017}. Then, we propose MSGAMP for joint activity detection and channel estimation.

% Joint channel and user activity estimation
% -------------------------------------------------------------------------
\section{Joint channel and user activity estimation}
\label{sec:jointAUDCD}

\begin{figure}[t]
            \centering
            \includegraphics[scale=0.9]{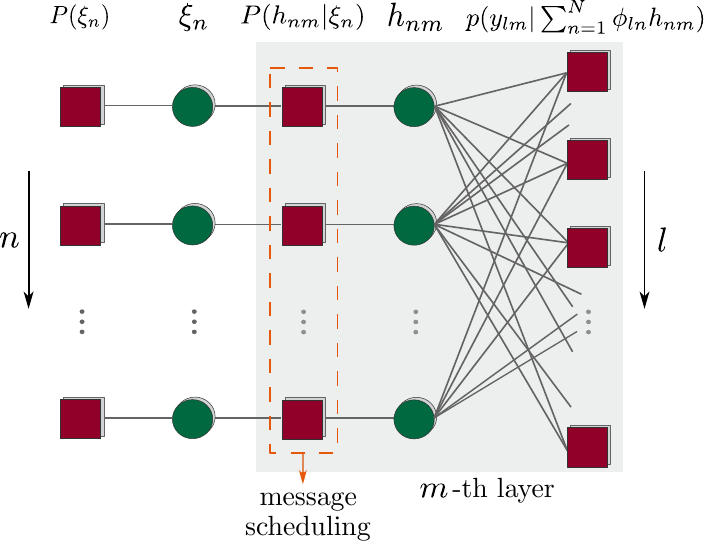}
            \caption{The factor graph pf joint distribution $p(\mathbf{H},\mathbf{Y},\boldsymbol{\xi})$ where cubes denote factor nodes and spheres variable nodes.}
            \label{fig:fg}%
        \end{figure}

       In this section we present the factor graph approach and depict the message-scheduling schemes. As described in Fig.~\ref{fig:fg}, given the independence of devices, we can write the prior distribution of $\mathbf{h}_m$ is $p\left(\bh_{m}|\xi\right) = \prod_{n=1}^N p\left(h_{nm}| \xi_n\right)$ and the likelihood, as given by
       \vspace{-5pt}
       \begin{align}
            p\left(\mathbf{Y}|\mathbf{Z}\right) &= \prod_{l=1}^{L}\prod_{m=1}^{M} p\left(y_{lm}|z_{lm}\right), \text{ where } \mathbf{Z} = \boldsymbol{\Phi}\,\mathbf{H}.
       \end{align}

       As the goal of GAMP, and consequently of HyGAMP, is to approximate the marginal posterior density by a product of the prior and a Gaussian distribution, the minimum mean squared error (MMSE) estimate of $h_{nm}$, $\hat{h}_{nm} = \mathbb{E}_{h_{nm}|\mathbf{y}}\left[h_{nm}\right] \forall\, n, m$, is given by $p\left(h_{nm}|\mathbf{Y}\right) = \int p\left(\mathbf{H}, \boldsymbol{\xi}|\mathbf{Y}\right)\, \text{d}\mathbf{\boldsymbol{\xi}}\, \text{d}\mathbf{H}_{\backslash nm},$. $\bH_{\backslash nm}$ denotes all elements except $h_{nm}$ and $p\left(\bH, \bxi|\bY\right)$ denotes the posterior distribution, given by the Bayes' rules
        \begin{align}\label{eq:post_dist}
           p\left(\bH, \bxi|\bY\right) =& \left[{1}/{p\left(\bY\right)}\right]\, p\left(\bY|\bH,\bxi\right) p\left(\bH | \bxi\right) p\left(\bxi\right)\\ \nonumber
           =& \left[{1}/{p\left(\bY\right)}\right]\, \left[\prod_{l=1}^{L} \prod_{m=1}^{M} p\left(y_{lm} |z_{lm}\right)\right]\\ \nonumber
           & \times \left[\prod_{n=1}^{N} \prod_{m=1}^{M} P\left(h_{nm}|\xi_n\right)\right] \left[\prod_{n=1}^{N} P\left(\xi_n\right)\right],
        \end{align}

        \noindent where $P\left(h_{nm}|\xi_n\right)$ is the conditional density for the random variable that $h_{nm} | \bxi \sim \delta({h_{nm}})$ if $(\xi_n =0)$ and $\mathcal{C}\mathcal{N}\left(h_{nm} | 0,\beta_n\right)$ otherwise.
        % Factor graph figure

% Factor graph approach --------------------------------------
\subsection{Factor graph approach}

        In order to marginalize the problem, Fig.~\ref{fig:fg} shows a factor graph (FG) that represents the factorization of~(\ref{eq:post_dist}). Spheres symbolizes variable nodes $\xi_n$ and $h_{nm}$ while factor nodes are symbolized as cubes. Factor nodes are the density functions, prior and likelihood. The approach that will be presented in what follows is inspired by work on low-density parity-check (LDPC) codes \cite{bfpeg,dopeg,memd} and message passing algorithms \cite{CasadoTCom2010,vfap,aaidd,kaids,dynmtc}.

        Since in our work the pilot matrix $\boldsymbol{\Phi}$ is a dense matrix, the FG in Fig.~\ref{fig:fg} is fully connected. The idea to easily compute the messages is to approximate them by prototype functions that resemble Gaussian density functions which can be described by two parameters only. This makes it possible to iteratively approximate, for a FG with cycles as in Fig.~\ref{fig:fg}, the marginal posteriors passing messages between different nodes. So, we can define the messages from $p\left(y_{lm} \big| \cdot\right)$ to $h_{nm}$ and to the opposite direction as
        \vspace{-10pt}
        \begin{flalign}\label{eq:mess1}
            \nu^{(i)}_{n\leftarrow lm}\left(h_{nm}\right) \propto&  \int p\left(y_{lm} \big| \sum_{k=1}^{N} \phi_{lk}\, h_{km}\right) \\ \nonumber
            & \times \prod_{j\neq n}^N \nu^{(i)}_{j \rightarrow lm}\left(h_{jm}\right)\, \text{d}h_{jm} \\\label{eq:mess2}
            \nu^{(i+1)}_{n\rightarrow lm}\left(h_{nm}\right) \propto&  \, \nu^{(i)}_{n\rightarrow nm}\left(h_{nm}\right)
            \prod_{k\neq m}^M \nu^{(i)}_{n \leftarrow lk}\left(x_{nk}\right)
        \end{flalign}
        \noindent and, considering $\propto$ as proportional, the messages from $P\left(h_{nm} | \xi_n\right)$ to $h_{nm}$ and to the opposite direction  are given by
        \vspace{-10pt}
        \begin{flalign}\label{eq:mess3}
            \nu^{(i)}_{n\leftarrow nm}\left(h_{nm}\right) \propto&  \prod_{k=1}^{M} \nu^{(i)}_{n\rightarrow lk}\left(h_{nm}\right),  \\\label{eq:mess4}
            \nu^{(i+1)}_{n\rightarrow nm}\left(h_{nm}\right) \propto &\, \int p\left(h_{nm} | \xi_{n}\right) \nu^{(i)}_{n\rightarrow nm}\left(\xi_n\right)\, \text{d}\xi_{n}.
        \end{flalign}

    Thus, the belief distribution that provides an approximation to marginal posterior distribution $p\left(h_{nm} | \bY, \bxi\right)$ is given by
    \begin{equation}\label{eq:Bdist}
        \nu^{(i+1)}_{nm}\left(h_{nm}\right) = \frac{\nu^{(i)}_{n\rightarrow nm} \prod_{s=1}^{M} \nu^{(i)}_{n\leftarrow ls}\left(h_{ns}\right)}{\int \nu^{(i)}_{n\rightarrow nm} \prod_{s=1}^{M} \nu^{(i)}_{n\leftarrow ls}\left(h_{ns}\right)\, \text{d}h_{nm}},
    \end{equation}
    \noindent where, defining
    \begin{equation}
    \resizebox{0.85\hsize}{!}{
        \hspace{-11pt}$p\left(h \big| \hat{r}_{nm}^{(i)},Q^{r(i)}_{nm};\hat{\rho}^{(i)}_{n}\right) \triangleq \frac{p\left(h; \hat{\rho}^{(i)}_n\right)\mathcal{C}\mathcal{N}\left(h| \hat{r}_{nm}^{(i)},Q^{r(i)}_{nm}\right)}{\int p\left(h; \hat{\rho}^{(i)}_n\right)\mathcal{C}\mathcal{N}\left(h| \hat{r}_{nm}^{(i)},Q^{r(i)}_{nm}\right)\,\text{d}h}$}
    \end{equation}
    \begin{equation}
    \resizebox{0.85\hsize}{!}{
        \hspace{-11pt}$p\left(z \big| p_{lm}^{(i)},Q^{p(i)}_{lm}\right) \triangleq \frac{p\left(y_{lm}|z^{(i)}_{lm}\right)\mathcal{C}\mathcal{N}\left(z | p_{lm}^{(i)},Q^{p(i)}_{lm}\right)}{\int p\left(y_{lm}|z^{(i)}_{lm}\right)\mathcal{C}\mathcal{N}\left(z | p_{lm}^{(i)},Q^{p(i)}_{lm}\right)\,\text{d}z}$,}
    \end{equation}

    \noindent we can compute $\mathbb{E}\left[\nu^{(i+1)}_{nm}\left(h_{nm}\right)\right] = \hat{h}^{(i+1)}_{nm}$ and $\text{Var}\left[\nu^{(i+1)}_{nm}\left(h_{nm}\right)\right] = Q^{h(i+1)}_{nm}$.

    Specifically, each iteration of the Algorithm 1 has three stages. The first stage, labelled as ``GAMP approximation'' contains the updates of the GAMP based on expectation propagation algorithm, which treats the components $h_{n}$ as independent with the probability of being active $\hat{\rho}_{n}$. Drawing inspiration from~\cite{AhnTCom2019, PSchniter2011, QZou2020}, we incorporated the EP in the process of LBP to relaxed belief propagation and then to GAMP. At iteration $i$, the MSGAMP algorithm produces estimates $\hat{\mathbf{h}}^{(i)}$ and $\hat{\mathbf{z}}^{(i)}$ of the vectors $\mathbf{h}$ and $\mathbf{z}$, as other intermediate vectors, $\hat{\mathbf{p}}^{(i)}$, $\hat{\mathbf{r}}^{(i)}$ and $\hat{\mathbf{s}}^{(i)}$. Covariance matrices like $\mathbf{Q}^{h(i)}$ and $\mathbf{Q}^{z(i)}$ are also produced.
    To be succinct, in order to reduce the complexity of $O\left(LNM\right)$ to $O\left(NM\right)$, the message in~(\ref{eq:mess1}) is firstly mapped to Gaussian distribution based on the central limit theorem and Taylor expansions. So, $\nu^{(i)}_{n\leftarrow lm}\left(h_{nm}\right)$ is updated by the Gaussian reproduction property. Following the same procedure in the messages of~(\ref{eq:mess2}), (\ref{eq:mess3}) and (\ref{eq:mess4}), relaxed BP is obtained by the combination of the approximated messages. Since many of these messages only differ slightly from each other, in order to fill up those differences, new variables are produced and ignoring the infinitesimals, the GAMP based on EP is obtained.

    The second stage of Algorithm 1, labelled as ``sparsity-rate update'', refers to the LBP part of the FG in Fig.~\ref{fig:fg} and updates the estimates of each probability of being active $\hat{\rho}_{nm}$. Computed using Gaussian approximations of likelihood functions, these estimates are then used to define the message scheduling proposed in this work. The messages in the ``sparsity-rate update'' stage are given by
    \vspace{-5pt}
        \begin{align}\label{eq:LBP1}
            \nu^{(i)}_{n \leftarrow nm}\left(\xi_n\right) \propto& \int p\left(h|\xi_n\right) \nu^{(i)}_{n \leftarrow nm}\left(h\right)\, \text{d}{h}, \\ \label{eq:LBP2}
            \nu^{(i)}_{n \rightarrow nm}\left(\xi_n\right) \propto&\, P\left(\xi_n\right) \prod_{k \neq m}^M \nu^{(i)}_{n \leftarrow nk}\left(\xi_n\right),
        \end{align}
    \noindent where (10) refers to the message from $P\left(h_{nm}| \xi_n \right)$ to $\xi_n$ while (\ref{eq:LBP2}) denotes the message in opposite direction and each belief at $\xi_n$ is given by $\nu^{(i)}_n\left(\xi_n\right) \propto P(\xi_n)\prod_{m=1}^M \nu^{i}_{n\leftarrow nm}(\xi_n)$.

    The message in (\ref{eq:LBP1}) can be approximated as a likelihood function, $\nu^{(i)}_{n \leftarrow nm}\left(\xi_n\right) = \mathcal{C}\mathcal{N}\left(h_{nm}|\hat{r}_{nm}^{(i)}, Q^{r(i)}_{nm} \right)$ and, applying the Gaussian reproduction property~\cite{QZou2020} thus enabling us to define
    \vspace{-5pt}
    \begin{align}\nonumber
        \text{LLR}_{n\leftarrow nm}^{(i)} =& \log \frac{\nu^{(i)}_{n \leftarrow nm}\left(\xi_n = 1\right)}{\nu^{(i)}_{n \leftarrow nm}\left(\xi_n = 0\right)}\\ \label{eq:LLRnleftnm}
        =& \log \frac{\mathcal{C}\mathcal{N}\left(0 \big| \hat{r}_{nm}^{(i)}, Q^{r(i)}_{nm} + \beta_n\right)}{\mathcal{C}\mathcal{N}\left(0 \big| \hat{r}_{nm}^{(i)}, Q^{r(i)}_{nm}\right)}.
    \end{align}
    Similarly to~(\ref{eq:LLRnleftnm}), we have $\text{LLR}_{n} \triangleq \log \frac{\nu^{(i)}_{n}\left(\xi_n = 1\right)}{\nu^{(i)}_{n}\left(\xi_n = 0\right)}$ and $\text{LLR}_{n\rightarrow nm}^{(i)} \triangleq \log \frac{\nu^{(i)}_{n \rightarrow nm}\left(\xi_n = 1\right)}{\nu^{(i)}_{n \rightarrow nm}\left(\xi_n = 0\right)}$. Substituting~(\ref{eq:LLRnleftnm}) in~(\ref{eq:LBP2}) and in each belief, $\text{LLR}_{n\rightarrow nm}^{(i)}$ is given by
    \vspace{-5pt}
    \begin{equation}
        \text{LLR}_{n\rightarrow nm}^{(i)} = \log \left(\frac{\rho_n}{1-\rho_n}\right) + \sum_{d\neq m}^M \text{LLR}_{n\leftarrow nd}^{(i)}.
    \end{equation}
   Thereby, the message in~(\ref{eq:mess4}) is
   \vspace{-5pt}
    \begin{align}
        \nu^{(i+1)}_{n\rightarrow lm}\left(h_{nm}\right) =&\, P\left(h_{nm}|\xi_n =1\right)\nu^{(i)}_{n\rightarrow nm}\left(\xi_n =1\right)\\ \nonumber
        & + P\left(h_{nm}|\xi_n =0\right)\nu^{(i)}_{n\rightarrow nm}\left(\xi_n =0\right)\\ \nonumber
        =&\, \hat{\rho}^{(i)}_{nm}\, \mathcal{C}\mathcal{N}\left(h_{nm}|0,\beta_n\right) + \left(1-\hat{\rho}^{(i)}_{nm}\right)\delta\left(h_{nm}\right),\\
        \triangleq&\, p\left(h_{nm}; \hat{\rho}^{(i)}_{nm}\right),
    \end{align}
   where
   \begin{equation}\label{eq:est_rho}
        \hat{\rho}^{(i)}_{nm} \triangleq \nu^{(i+1)}_{n\rightarrow lm}\left(\xi_n =1\right) = 1 - \frac{1}{1 + \exp\left(\text{LLR}_{n\leftarrow nm}^{(i)}\right)}.
   \end{equation}

    With the HyGAMP algorithm established in the context of mMTC message-scheduling techniques can be applied.
    \vspace{10pt}

% Algorithm pseudocode -----------------------------------
\input{alg.tex}

% Message-scheduling schemes ------------------------------
\subsection{Message-scheduling schemes}
    In this subsection, we propose three ordering schemes for designing the message scheduling of MSGAMP. RBP is an informed dynamic strategy that updates messages according to an ordering metric called the residual. A residual is the norm (defined over the message space) of the difference between the values of a message before and after an update. In our scheme, we define the residual with the beliefs depicted in~(\ref{eq:Bdist}). Thus, the residual for the belief distribution at $h_{nm}$, is given by
    \begin{equation}\label{eq:Res}
        \text{Res}\left(\nu_{nm}\left(h_{nm}\right)\right) = \big|\big|\nu^{(i+1)}_{nm}\left(h_{nm}\right) - \nu^{(i)}_{nm}\left(h_{nm}\right)\big|\big|.
    \end{equation}

    The intuitive justification of this method is that as the factor graph approach converges, the differences between the messages before and after an update diminish. Therefore, if a message has a large residual, it means that it is located in a part of the graph that has not converged yet. Thus, propagating that message first should speed up the convergence. Based on this criterion, the three methods to be presented in this work update the messages and the stopping criterion. Using the residual values computed in~(\ref{eq:Res}), we compute the set $\bS^{(i)}$ of messages to be updated in the next iteration. Taking advantage of a massive MIMO scenario where mMTC is inserted, MSGAMP uses the estimates of each BS antenna to refine the activity detection, as $\hat{\rho}_{n}^{(i)} = \sum_{m=1}^M \hat{\rho}^{(i)}_{nm}/M$, where if $\hat{\rho}_{n}^{(i)}$  is higher than a threshold, the device is considered active. Thus, MSGAMP proceeds until $i$ surpasses the maximum number of iterations $I$ or $\left(\text{tol}/M < 10^{-4}\right)$, where tol is given by
    \begin{equation}\label{eq:StopCrit}
        \text{tol} = \sum_{m=1}^M \frac{\|\hat{{\bh}}^
        {(i)}_{m} - \hat{{\bh}}^
        {(i-1)}_{m}\|}{\|\hat{{\bh}}^
        {(i)}_{m}\|}.
    \end{equation}

% MSGAMP-RBP ------------------------------
\subsubsection{MSGAMP-RBP}
  The Message Scheduling based on Residual Belief Propagation (MSGAMP-RBP) is a dynamic scheduling strategy that updates only the set of messages of the node that has the largest residual, computed in~(\ref{eq:Res}), that is, $\|\bS^{(i)}\| =1$. The values of the messages that belong to other nodes are repeated. The residuals are computed at each new iteration, so that the focus of the updates is only on the messages from the node with the largest residue.

% MSGAMP-GRBP ------------------------------
\subsubsection{MSGAMP-GRBP}
    In this strategy, we have a message scheduling inside the group of messages that belongs to a set of nodes with the largest residues. In the Message Scheduling GAMP based on Group Residual Belief Propagation (MSGAMP-GRBP), we update only the messages of the group, in a sequentially manner, as described in Algorithm~2. The size of the group is defined by the activity detection, that is, the group is composed by the $|\hat{\boldsymbol{\rho}}^{(i)}|$ nodes with largest residuals.

    % Algorithm pseudocode | message scheduling based on group AUD
    \input{alg_GRBP.tex}

    When $i=2$, we have the first values of the residuals, thus enabling update of the set $\mathbf{S}^{(i)}$. In the next iteration, all messages that belong to $\mathbf{S}^{(i)}$, except for $s^{(i)}_1$ will be updated. Then, the index that refers to the group of messages that had been updated is removed of $\mathbf{S}^{(i)}$ as in
    \begin{equation}
        \mathbf{S}^{(i)} = \left[s^{(i-1)}_2, \dots, s^{(i-1)}_{|\bS^{(i-1)}|}\right].
    \end{equation}
    Therefore, we exclude a group of messages that belong to a specific device to be updated, one at a time. When $\bS^{(i)}$ is empty, a new update of the set $\bS^{(i)}$ is performed. In a nutshell, we reduce the set $\bS^{(i)}$ that is updated in parallel, until there is no message to update.

% MSGAMP-GRBPp ------------------------------
\subsubsection{MSGAMP-GRBPp}
    Since the last message scheduling techniques do not update the group of messages outside the set $\bS^{(i)}$, they could suffer if the residual computation is not properly done. In order to address such possible errors, when a new set $\bS^{(i)}$ is computed, the Message Scheduling GAMP based on Group Residual Belief Propagation with Parallel update (MSGAMP-GRBPp) modifies the messages that do not belong to the set, as given by
    \begin{equation}
    \left\{
        \begin{array}{rr}
            \bS^{(i)} = \left[1, \dots, N\right], &  |\bS^{(i-1)}| = 0,\\
            \text{proceed as in MSGAMP-GRBP}, & \text{otherwise.}
        \end{array}\right.
    \end{equation}
    After the first update of the set $\bS^{(i)}$ is made, MSGAMP-GRBPp proceeds as MSGAMP-GRBP. When $\bS^{(i)}$ is empty, all messages are updated and a new activity detection is performed, as a new $\bS^{(i)}$.

    Since in future mobile communication systems it is expected up to $300,000$ devices per cell~\cite{3GPPTR36888} and the sporadic transmission pattern of each device, the gain in scheduling schemes is relevant. Unlike the well-known parallel message update of HyGAMP that, in each iteration, $O(MN)$ messages must be computed, in our schemes we need only $O(M|\bS^{(i)}|)$. With the residual computation, the activity detection procedure and stopping criterion defined, we present the first message scheduling scheme, MSGAMP-RBP.

% Simulation results
% -------------------------------------------------------------------------
\section{Simulation results}
\label{sec:sim}
\begin{figure}[t]
        \begin{minipage}[b]{1\linewidth}
          \centering
          \centerline{\includegraphics[scale=0.95]{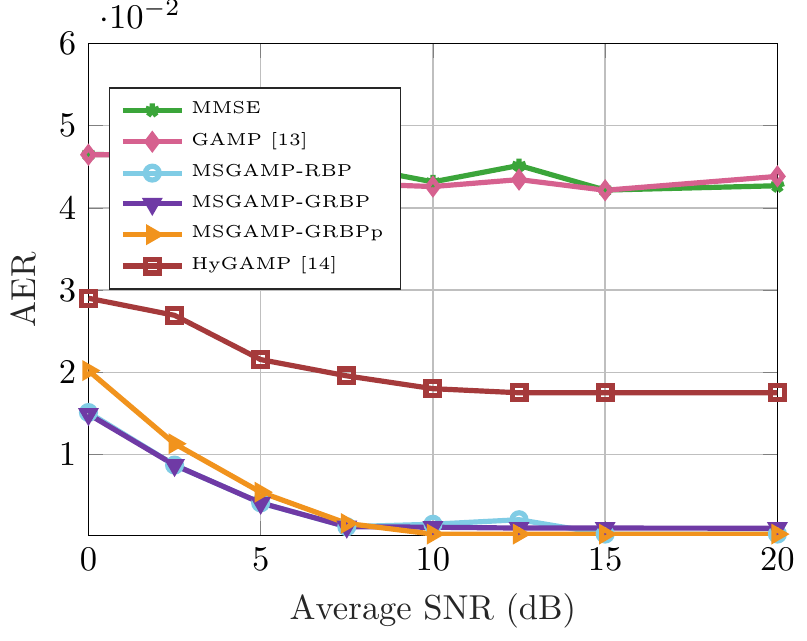}}
          \centerline{\footnotesize (a) Activity error rate results.}\medskip
        \end{minipage}

        \begin{minipage}[b]{1\linewidth}
          \centering
          \centerline{\includegraphics[scale=0.95]{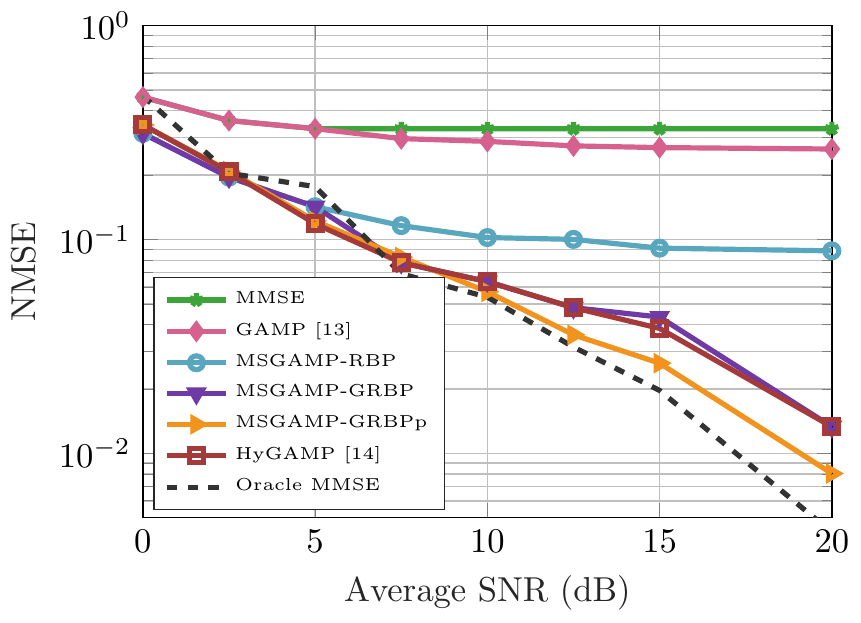}}
          \centerline{\footnotesize (b) Normalized mean squared error of the active devices.}\medskip
        \end{minipage}
    \caption{Performance in terms of AER and NMSE versus average SNR. The NMSE considered only the active devices.}
    \label{fig:sim}%
\end{figure}

\begin{figure}[t]
        \begin{minipage}[b]{0.5\linewidth}
          \centering
          \centerline{\includegraphics{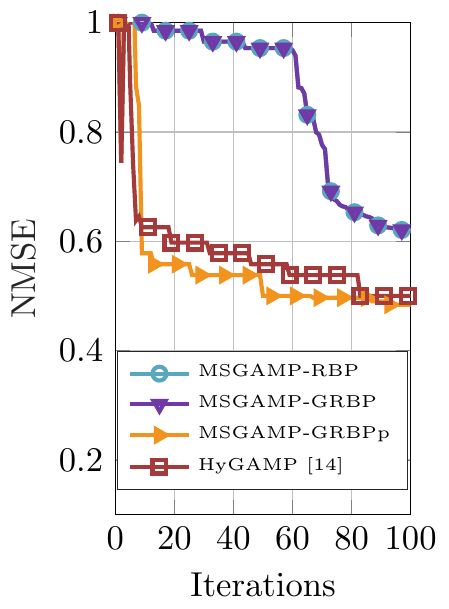}}
          \centerline{(a) SNR = 5 dB}\medskip
        \end{minipage}
        \begin{minipage}[b]{0.48\linewidth}
          \centering
          \centerline{\includegraphics{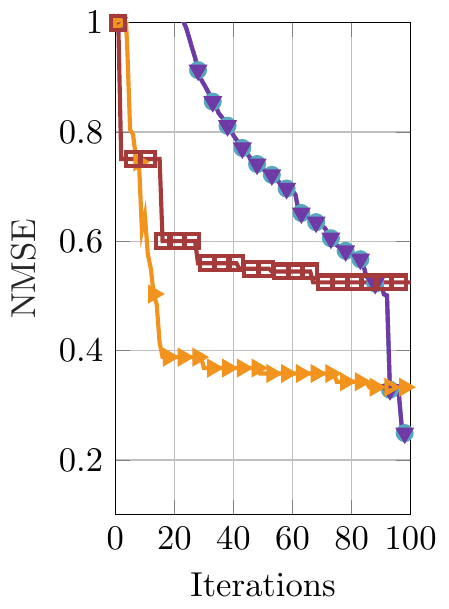}}
          \centerline{(b) SNR = 10 dB}\medskip
        \end{minipage}

        \begin{minipage}[b]{0.5\linewidth}
          \centering
          \centerline{\includegraphics{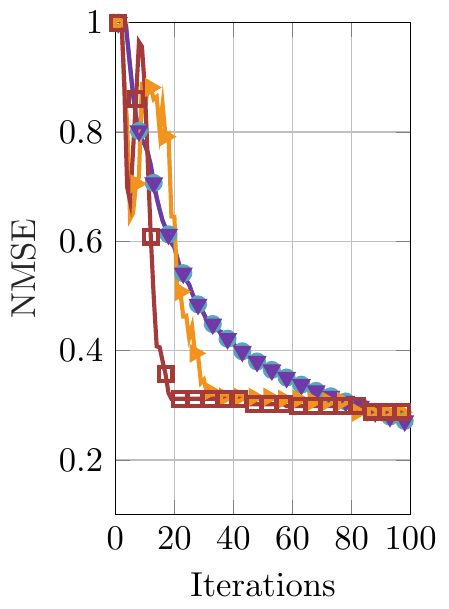}}
          \centerline{(c) SNR = 15 dB}\medskip
        \end{minipage}%
        \begin{minipage}[b]{0.5\linewidth}
          \centering
          \centerline{\includegraphics{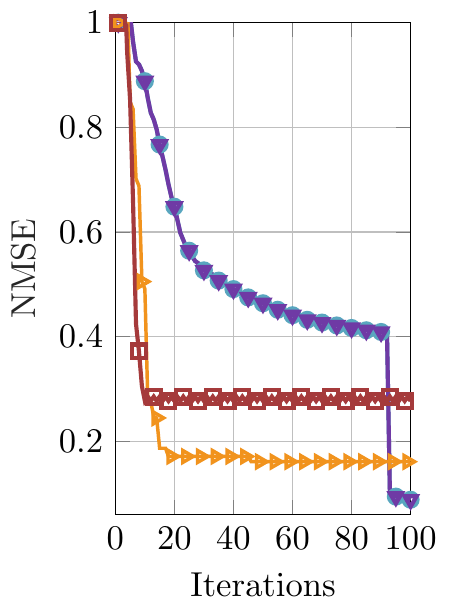}}
          \centerline{(d) SNR = 20 dB}\medskip
        \end{minipage}%
    \caption{Convergence rate of MSGAMP-type techniques and HyGAMP. The NMSE considered all channel gains, not only the ones of active devices.}
    \label{fig:Conv}%
\end{figure}

    In order to verify the performance of the proposed MSGAMP schemes, as described in Section~\ref{sec:sys_mod},
    we simulate an mMTC system with $N=128, M=32$ and $L=64$. The threshold to detect the activity of devices considered is 0.9, the average SNR is set to $1/\sigma_w^2$, while the activity probabilities $p_n$ are drawn uniformly at random in $\left[0.01,0.05\right]$. The variations of MSGAMP are compared to the well-known minimum mean squared error (MMSE), generalized approximate message passing (GAMP)~\cite{Rangan2012} and the state-of-the-art algorithm, HyGAMP~\cite{Rangan2017}. An MMSE detector with perfect activity knowledge (oracle MMSE) is used as a lower bound. Figs.~\ref{fig:sim}a and~\ref{fig:sim}b show results of AER and NMSE. One can see that the use of the BS antennas in order to refine the activity detection improved the AER performance since the proposed MSGAMP-type techniques outperform HyGAMP. Regarding NMSE, Fig.~\ref{fig:sim}b shows that the message scheduling schemes have a competitive performance, where MSGAMP-GRBP achieves the efficiency of HyGAMP while MSGAMP-GRBPp outperforms both, requiring less computational cost. Fig.~\ref{fig:Conv} depicts the convergence rate of MSGAMP-type techniques and HyGAMP. One can notice that for different values of SNR, MSGAMP-type techniques converge faster and to lower values of NMSE than HyGAMP.

% AUD and NMSE results

% Convergence results

% Conclusion
% -------------------------------------------------------------------------
\section{Conclusion}
\label{sec:conclusion}
    In this paper we have presented MSGAMP-type techniques that perform joint activity detection and channel estimation for mMTC. Exploiting the BS antennas, we developed three scheduling techniques for MSGAMP that update the messages based on the residual belief propagation. The results indicate that MSGAMP-type techniques outperform other solutions in terms of NMSE and AER, while requiring a small number of iterations for convergence.

\section*{Acknowledgment}
    The authors would like to thank the Conselho Nacional de Desenvolvimento Cient\'{i}fico e Tecnol\'{o}gico (CNPq), Fundação de Amparo à Pesquisa no Rio de Janeiro (FAPERJ) and the Coordena\c{c}\~{a}o de Aperfei\c{c}oamento de Pessoal de N\'{i}vel Superior - Brasil (CAPES) by funding.

%\vfill\pagebreak

% References
% -------------------------------------------------------------------------
\bibliographystyle{IEEEbib}
\bibliography{refs}

\end{document}

%% file: alg.tex
% Algorithm pseudocode
% -------------------------------------------------------------------------
\vspace*{-\baselineskip}
\begin{table}[t]
	%\vspace{-5pt}
%\normalsize
%\small
\footnotesize
%\scriptsize
	\begin{center}  
		\begin{tabular}{p{8cm}l} \\ \hline
		{\footnotesize\textbf{Algorithm 1} MSGAMP-AUD for mMTC} \\ \hline
            \textbf{initialize}\\
            \hspace{4.5pt}$1$:\hspace{5pt} $i=1$, $\hat{{s}}_{lm}^{(0)} = 0$, $\hat{{r}}_{nm}^{(0)} = 0$, $Q^{r(0)}_{nm} = 1$, $\hat{\rho}^{(0)}_{nm} = \rho_{n}$,\\
            \hspace{15pt} $\bS^{(0)} = \left[1, \dots, N\right]$ \\
            % repeat -----
            \textbf{repeat}\\
            \hspace{5pt} \% \textit{GAMP approximation}\\
            % FOR M -----            
            \hspace{4.5pt}$2$:\hspace{5pt} \textbf{for} $\left(n= 1, \dots, |\bS^{(i-1)}|\right) \forall n \in \bS^{(i-1)} $ \\
            % FOR N -----                        
            \hspace{4.5pt}$3$:\hspace{15pt} \textbf{for} $\left(m= 1, \dots, M\right)$ \\
            \hspace{4.5pt}$4$:\hspace{25pt} $\hat{h}_{nm}^{(i)} = \mathbb{E}\left[\mathcal{X}_{nm}\big|\hat{r}_{nm}^{(i-1)},Q^{r(i-1)}_{nm};\hat{\rho}^{(i-1)}_{n}\right]$ \\[2pt]
            \hspace{4.5pt}$5$:\hspace{25pt} ${Q^{h(i)}_{nm}} = \text{Var}\left[\mathcal{X}_{nm}\big|\hat{r}^{(i-1)}_{nm},Q^{r(i-1)}_{nm};\hat{\rho}^{(i-1)}_{n}\right]$ \\[2pt]            
            % FOR Tau -----                        
            \hspace{4.5pt}$6$:\hspace{25pt} \textbf{for} $\left(l= 1, \dots L \right)$ \\[2pt] 
            %$6$:\hspace{25pt} ${\hat{z}^{(i)}_{tm}} = \sum_{n=1}^N \Phi_{tn} \hat{h}_{nm}^{(i)}$; \\                                    
            \hspace{4.5pt}$7$:\hspace{35pt} $Q^{p(i)}_{tm} = \sum_{n=1}^N |\Phi_{tn}|^2 Q^{h(i)}_{nm}$ \\[2pt] 
            \hspace{4.5pt}$8$:\hspace{35pt} $p^{(i)}_{tm} = \sum_{n=1}^N \Phi_{tn}\, \hat{h}_{nm}^{(i)} - Q^{p(i)}_{tm} \hat{s}_{tm}^{(i-1)}$ \\[2pt] 
            \hspace{4.5pt}$9$:\hspace{35pt} $\tilde{z}^{(i)}_{tm} = \left(y_{lm}\, Q^{p(i)}_{tm} + \sigma^2_w p^{(i)}_{tm}\right)\big/\left(Q^{p(i)}_{tm} + \sigma^2_w\right)$ \\[2pt]  
            $10$:\hspace{35pt} $Q^{z(i)}_{tm} = \left(\sigma_w^2 Q^{p(i)}_{tm}\right)\big/\left(Q^{p(i)}_{tm} + \sigma^2_w\right)$ \\[2pt]              
            $11$:\hspace{35pt} $\hat{s}^{(i)}_{tm} = \left(\tilde{z}^{(i)}_{tm}-p^{(i)}_{tm}\right)\big/Q^{p(i)}_{tm}$ \\[2pt] 
            $12$:\hspace{35pt} $Q^{s(i)}_{tm} = Q^{-p(i)}_{tm} \left(1 - Q^{z(i)}_{tm}/Q^{p(i)}_{tm}\right)$  \\[2pt]             
            $13$:\hspace{22pt} \textbf{end for} \\[2pt] % end
            $14$:\hspace{25pt} $Q^{-r(i)}_{nm} = \sum_{l=1}^L |\Phi_{tn}|^2 Q^{s(i)}_{tm}$ \\[2pt]  
            $15$:\hspace{25pt} $r^{(i)}_{nm} = \hat{h}_{nm}^{(i)} + Q^{r(i)}_{nm} \sum_{l=1}^{L} \Phi^\ast_{tn} \hat{s}^{(i)}_{tm}$ \\[2pt]
            \hspace{5pt} \% \textit{Sparsity-rate update}\\   
            $16$:\hspace{25pt} $\text{LLR}_{n\leftarrow nm}^{(i)} = \log \frac{\mathcal{C}\mathcal{N}\left(0 \big| \hat{r}_{nm}^{(i)}, Q^{r(i)}_{nm} + \beta_n\right)}{\mathcal{C}\mathcal{N}\left(0 \big| \hat{r}_{nm}^{(i)}, Q^{r(i)}_{nm}\right)}$ \\[2pt]
            $17$:\hspace{25pt} $\text{LLR}_{n\rightarrow nm}^{(i)} = \log \left(\nicefrac{\rho_n}{1-\rho_n}\right) + \sum_{d\neq m}^M \text{LLR}_{n\leftarrow nd}^{(i)}$\\[2pt]  
            $18$:\hspace{25pt} $\hat{\rho}_{nm}^{(i)} = 1 - \nicefrac{1}{1 + \exp{\left(\text{LLR}_{n\rightarrow nm}^{(i)}\right)}}$\\[1pt]                          
            $19$:\hspace{12pt} \textbf{end for} \\[2pt]% end             
            $20$:\hspace{2pt} \textbf{end for}  \\[2pt]% end
            \hspace{5pt} \% \textit{Message scheduling update}\\  
            $21$:\hspace{5pt} $\hat{\rho}_{n}^{(i)} = \sum_{m=1}^{M}
            \hat{\rho}_{nm}^{(i)}\big/M$  \\[2pt]
            $22$:\hspace{5pt} $\bS^{(i)} = \text{update}\left[\bS^{(i-1)}\right]$ \% \textit{chosen RBP message scheduling type}  \\[2pt]
            $23$:\hspace{5pt} $\text{tol} = \left(\nicefrac{1}{M}\right) \sum_{m=1}^M \|\hat{\underline{\bh}}^{(i)}_{m} -\hat{\underline{\bh}}^{(i-1)}_{m}\|\big/\,\|\hat{\underline{\bh}}^{(i)}_{m}\|$  \\[2pt]   
            $24$:\hspace{5pt} $i = i +1$ \\[2pt]              
            \textbf{until} $\left(i > I \text{ or tol} < 10^{-4} \right)$ \\[1ex]     \hline   
		\end{tabular}
	\end{center}
	\vspace{-15pt}
\end{table} 

%% file: alg_GRBP.tex
% Algorithm pseudocode
% -------------------------------------------------------------------------
\begin{table}[t]
	%\vspace{-3pt}%\normalsize
%\small
\label{tab:algGRBP}
\scriptsize
	\begin{center}  
		\begin{tabular}{p{8cm}l} \\ \hline
		{\footnotesize\textbf{Algorithm 2:} Message scheduling based on GRBP}\\ \hline
            \textbf{initialize}\\
            \hspace{4.5pt}$1$:\hspace{5pt} $l=1$ \\  		
            % IF -----                          
            \hspace{4.5pt}$2$:\hspace{5pt} \textbf{if} $\left(|S^{(i)}| = 0\right)$ \\
            % FOR N -----                        
            \hspace{4.5pt}$1$:\hspace{15pt} \textbf{for} $\left(n= 1, \dots, N\right)$ \\            
            % IF -----              
            \hspace{4.5pt}$4$:\hspace{25pt} \textbf{if} $\left(\text{Res}\left(\nu_{nm}\left(h_{nm}\right)\right) > \textit{the first other }|\hat{\boldsymbol{\rho}}^{(i)}| \textit{residuals}\right)$ \\
            \hspace{4.5pt}$5$:\hspace{35pt} $S^{(i+1)}_l = n$ \\            
            \hspace{4.5pt}$6$:\hspace{35pt} $l = l+1$\\            
            \hspace{4.5pt}$7$:\hspace{25pt} \textbf{end if}\\ 
            % END N -----                
            \hspace{4.5pt}$8$:\hspace{15pt} \textbf{end for}\\                   
            \hspace{4.5pt}$9$:\hspace{5pt} \textbf{else}\\	
            % END IF -----                          
            \hspace{15pt} \% Remove the first element of $S^{(i)}$, that is, \\
            $10$:\hspace{15pt} $S^{(i+1)} = \left[S^{(i)}_2, \dots, S^{(i)}_{|S^{(i)}|}\right]$ \\
            $11$:\hspace{5pt} \textbf{end if}\\	            
            % FOR m -----              
            $12$:\hspace{5pt} $\forall \left(n \notin S^{(i+1)}\right)$, \\		
            $13$:\hspace{5pt} \textbf{for} $\left(m= 1, \dots, M\right)$ \\	 
            $14$:\hspace{15pt} $\hat{h}_{nm}^{(i+1)} = \hat{h}_{nm}^{(i)}$;  $Q^{h(i+1)}_{nm} = Q^{h(i)}_{nm}$; \\
            $15$:\hspace{15pt} $\hat{r}_{nm}^{(i+1)} = \hat{r}_{nm}^{(i)}$; $Q^{r(i+1)}_{nm} = Q^{r(i)}_{nm}$;\\           
            % END N -----                                                  
            $17$:\hspace{5pt} \textbf{end for}\\[1ex]     \hline   
		\end{tabular}
	\end{center}
	\vspace{-15pt}
\end{table} 
%\vspace{-4mm} 